\documentclass[twocolumn,english,superscriptaddress,prl]{revtex4}
\usepackage{subfigure}
\usepackage{amsmath}
\usepackage{graphicx}
\usepackage{xcolor}

\begin{document}

\title{Photonic Feshbach Resonance}
\author{D. Z. Xu}
\affiliation{Institute of Theoretical Physics, The Chinese Academy of Sciences,100190,
P.R. China}
\affiliation{Department of Modern Physics, University of Science and Technology of China,
Hefei, 230026, P.R. China}
\author{H. Ian}
\affiliation{Institute of Theoretical Physics, The Chinese Academy of Sciences,100190,
P.R. China}
\author{T. Shi}
\affiliation{Institute of Theoretical Physics, The Chinese Academy of Sciences,100190,
P.R. China}
\author{H. Dong}
\email{dhui@itp.ac.cn}
\affiliation{Institute of Theoretical Physics, The Chinese Academy of Sciences,100190,
P.R. China}
\author{C. P. Sun}
\email{suncp@itp.ac.cn}
\affiliation{Institute of Theoretical Physics, The Chinese Academy of Sciences,100190,
P.R. China}
\maketitle

Hermann Feshbach predicted fifty years ago~\cite{feshbach58} that
when two atomic nuclei are scattered within an open entrance
channel--- the state observable at infinity, they may enter an
intermediate closed channel
--- the locally bounded state of the nuclei. If the energy of a bound state
of in the closed channel is fine-tuned to match the relative kinetic
energy, then the open channel and the closed channel ``resonate'',
so that the scattering length becomes divergent~\cite{pethick02}. We
find that this so-called Feshbach resonance phenomenon not only
exists during the collisions of massive particles, but also emerges
during the coherent transport of massless particles, that is,
photons confined in the coupled resonator arrays~\cite{lzhou08}. We
implement the open and the closed channels inside a pair of such
arrays, linked by a separated cavity or a tunable qubit. When a
single photon is bounded inside the closed channel by setting the
relevant physical parameters appropriately, the vanishing
transmission appears to display this photonic Feshbach resonance.
The general construction can be implemented through various
experimentally feasible solid state systems, such as the couple
defected cavities in photonic crystals. The numerical simulation
based on finite-different time-domain(FDTD) method confirms our
conceive about physical implementation.

\bigskip

The phenomenon of Feshbach resonance has been found in many physical systems
over the years, such as the electron scattering of atoms~\cite{schulz73a}
and diatomic molecules~\cite{schulz73b}. More recently, the development of
laser cooling technologies has enabled the observation of low-energy
Feshbach resonance in ultra-cold atoms~\cite%
{courteille98,roberts98,vuletic99} and Bose-Einstein condensates (BEC)~\cite%
{ketterle98,timmermans99}. These experiments have helped verify the
simulation of various theoretical predictions of condensing phenomena in
solid state systems~\cite{dsjin08rev}. The latter in particular exemplifies
the resonance phenomenon as a means for adjusting inter-atomic coupling in
realizing various quantum phases ranging from BEC to BCS~\cite%
{d.s.jin02,d.s.jin08}.

On the other hand, since their first discovery by von Neumann and
Wigner~\cite{neumann29}, bound states have been studied in a general
continuum~\cite{friedrich85} and the emergence of a bounded energy
level has been verified by various
models~\cite{lee54,fano61,anderson61}, extracting from the simple
coupling of a discrete level with a continuum. Quasi-bound states
have been predicted in tight-binding fermionic quantum
wires~\cite{nakamura07} for localized fermions and in the optical
coupled resonator arrays~\cite{lzhou08,hdong08,zrgong08} for
confined photons. It has also found various applications in many
quantum optical devices~\cite{fan05,fan05_optic,fan07}, including
the single-photon transistor~\cite{lukin07}. Under this retrospect,
we ask if it is possible to control exactly when the photons become
bounded and unbounded through an external parameter, \emph{viz.} to
implement an optical version of Feshbach resonance.

\begin{figure}[tbp]
\includegraphics[bb=58bp 375bp 512bp 760bp,clip,width=8cm]{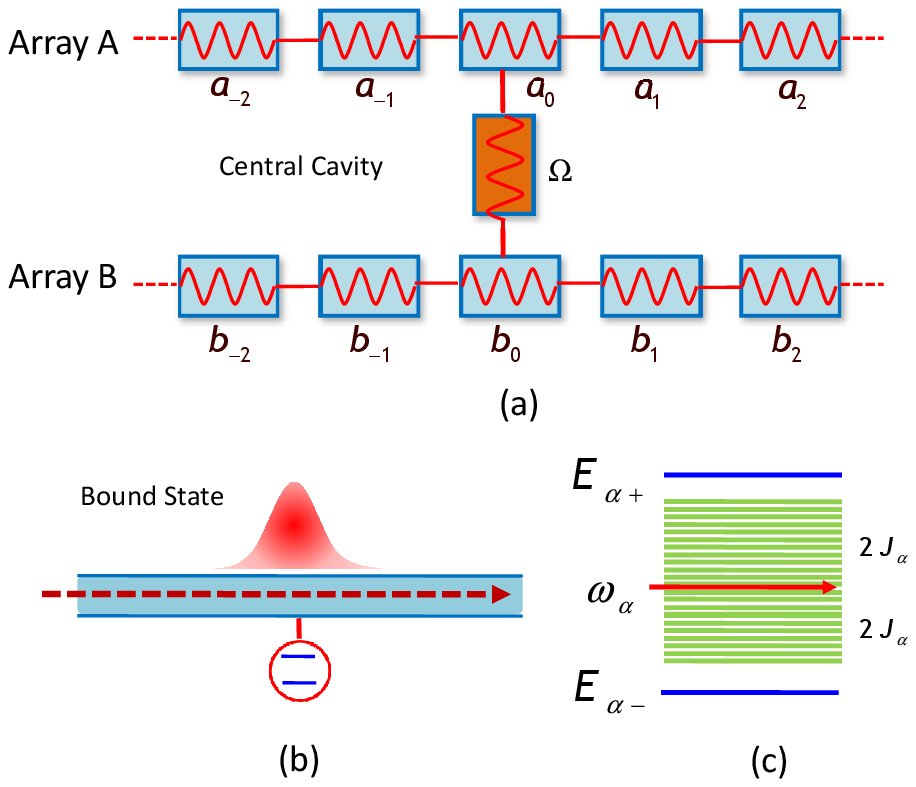}
\caption{The photonic  Feshbach resonance based on the coupled
resonator arrays: (a) schematic of the H-shape system consisting of
two coupled resonator arrays, array A and array B, connected by a
qubit controller of level spacing $\Omega $. Each of the cavity in
the two arrays is characterized by the cavity field mode it confines
and hereby indicated by the bosonic operator $a_{j}$ or $b_{j}$ and
indexed by the relative distance $j$ from the zeroth cavity with
which the qubit controller couples; (b) probability distribution of
a single photon in a coupled resonator array for a quasi-bound
state. When the level spacing $\Omega $ of the qubit controller
matches certain values, the probability amplitude would vanish at
the two ends of the array; (c) the energy state distribution for
array $\alpha $, $\alpha \in \left\{ A,B\right\} $, relative to the
common eigenfrequency $\omega _{\alpha }$ of the cavity fields: a
pair of a continuum and a discrete level above and below $\omega
_{\protect\alpha }$.} \label{fig:model}
\end{figure}

The desired resonance between the bound and the unbound states can
be found inside a pair of parallelly placed coupled resonator
arrays~\cite{lzhou08}, a series of consecutively placed optical
microcavities that entrap photons and allow photon-hopping from one
of the cavities to its closest neighbors at left and right. The two
arrays are connected by a  central cavity that acts as a quantum
controller and couples separately to one cavity in each of the
arrays, as shown in Fig.~\ref{fig:model}(a), forming an H-shape
system. We designate the upper array as array A with the Hamiltonian
\begin{equation}
H=\omega _{\mathrm{A}}\sum_{j}a_{j}^{\dagger }a_{j}+(J_{\mathrm{A}%
}\sum_{j}a_{j}^{\dagger }a_{j+1}+g_{\mathrm{A}}a_{0}c^{\dagger
}+\mathrm{h.c}), \label{eq:ham_array_A}
\end{equation}%
where the second summation describes the tight-binding hopping of
photons between neighboring cavities with hopping coefficient
$J_{\mathrm{A}}$ while the first one accounts for the static photon
occupations in the cavities. $a_{j}$ denotes the annihilation
operator of the bosonic mode for the $j$-th cavity field and we
assume that the mode frequency $\omega _{\mathrm{A}}$ of each cavity
field is identical. The last term is the interaction between the
central cavity $C$ of single mode frequency $\Omega $ and the zeroth
resonator of array A with coupling strength $g_{\mathrm{A}}$. $c$ is
corresponding annihilation bosonic operator for central cavity. The
lower array is designated as array B, the Hamiltonian
$H_{\mathrm{B}}$ for which is no different from
Eq.~(\ref{eq:ham_array_A}) except the change of the bosonic mode
operator to $b_{j}$,the hopping coefficient to $J_{\mathrm{B}}$, the
mode frequency to $\omega _{\mathrm{B}}$ and the coupling strength
to $g_{\text{B}}$. Then,the total model Hamiltonian reads
$H=H_{\mathrm{A}}+H_{\mathrm{B}}+\Omega c^{\dagger }c.$

One experimental available system of the model is based on photonic
crystal, which will be described in details below. We have to point
out that there is a drawback in our setup: the inter-cavity coupling
cannot be externally manipulated, once the photonic-crystal based
metamaterial is fabricated as an all-optical chip. However, for
single photon transferring, the role  of central cavity as that of a
qubit with  two levels  $\left\vert e\right\rangle =\left\vert
1\right\rangle$ and $\left\vert g\right\rangle =$ $\left\vert
0\right\rangle$, corresponding to the single photon state and vacuum
of the central cavity. This identification motivates us to use a
qubit controller  replacing the central cavity equivalently in the
single photon case. Therefore, the phenomenon predicted here can be
realized in the hybrid system of photonic crystal and quantum dots
with external controllable parameter. Our general construction can
also be implemented through the circuit QED system~\cite{wallraff04}
including two coupled superconducting transmission line cavity
linked by charge or flux qbit. The replacement of the central cavity
by a controllable qubit can overcome the drawback  mentioned above.

In the following , we  use the Jaynes-Cummings couplings
$g_{\alpha}c^{\dagger }\left\vert g\right\rangle \left\langle
e\right\vert +\mathrm{h.c}$ ($\alpha=A,B$) to modeling the central
cavity couplings in the single photon case. The Hilbert space  is
spanned by the tensor product $\left\{ \left\vert e\right\rangle
,\left\vert g\right\rangle \right\} \bigotimes_{j}\left\vert
n_{\mathrm{A},j}\right\rangle \bigotimes_{j}\left\vert
n_{\mathrm{B},j}\right\rangle $ where $\left\vert n_{\alpha
,j}\right\rangle $ denotes the state of array $\alpha $ with its $j
$-th cavity being occupied with $n_{\alpha ,j}$ number of photons.
When separated from the other and studied individually, each coupled
resonator array is described by the subsystem Hamiltonian $H_{\alpha
}$ and possesses two bound states, reminiscent that of the Feshbach
resonance. The states are the particular superposition of
eigenstates for the Hamiltonian $H_{\alpha }$, comprised by a subset
of the basis vectors described above. For array A, the state is
namely $\left\vert \varphi _{\mathrm{A}}\right\rangle
=\sum_{j}u_{\mathrm{A},g}(j)\left\vert g,1_{j},0\right\rangle
+u_{\mathrm{A},e}\left\vert e,0,0\right\rangle $ where only the
excited state of the central cavity and a single-photon excitation
in one of the cavities, as indicated by the $1_{j}$ symbol, are
included. The coefficients $u_{\mathrm{A},g}(j)$ and
$u_{\mathrm{A},e}$ in the equation constitute the spectrum of
probability distributions of these states. That of array B takes a
similar form with amplitudes $u_{\mathrm{B},g}(j)$ and
$u_{\mathrm{B},e}$.

To see whether there are bounded single-photon states within their
individual coupled resonator array, we can solve the discrete-coordinate
scattering equation associated with their corresponding eigenvalue $E$ for
the probability spectrum,
\begin{equation}
\lbrack E-\omega _{\alpha }-V_{\alpha }(E)]u_{\alpha ,g}(j)=-J_{\alpha
}[u_{\alpha ,g}(j+1)+u_{\alpha ,g}(j-1)]  \label{eq:scatter_eqn}
\end{equation}%
where the term $V_{\alpha }(E)=g_{\alpha }^{2}\delta _{j0}/(E-\Omega
)$ on the left hand side is contributed by the JC type interaction
between the central cavity and the coupled resonator array.
$V_{\alpha }(E)$ is a resonate potential that depends on the
eigenenergy $E$. In the continuous limit of the coordinate
$j\rightarrow x$, this term reduces to a $\delta $ -type potential
\begin{equation}
V_{\alpha }(E,x)=\frac{g_{\alpha }^{2}}{E-\Omega }\delta (x).
\label{eq:eff_pot}
\end{equation}%
The $\delta $-type potential forms a confining barrier to the
transportation of single photon in the coupled resonator array and
informs a bounded single photon within, similar to those in the
models proposed in Refs.~\cite{hdong08,zrgong08}.

It has a singularity at $E$ being equal to the level spacing $\Omega
$, leading to a quasi-plane-wave type solution~\cite{lzhou08} to
Eq.~(\ref{eq:scatter_eqn}), $u_{\alpha ,g}(j)=C_{\alpha }\exp
(-i\kappa _{\alpha }|j|) $ where $C_{\alpha }$ denotes a constant
and the wave number is complex, $\kappa _{\alpha }=\kappa _{\alpha
,R}-i\kappa _{\alpha ,I}$. The imaginary part $\kappa _{\alpha ,I}$
of the wave number can admit a positive value and for the non-zero
coupling $g_{\alpha }$, resulting in a decay of the probability
distribution of single-photon states over the discrete spatial
coordinate $j$. The vanishing probability amplitude towards the ends
of the arrays, i.e. along with $|j|\rightarrow \infty $,
demonstrates the existence of a bound state of a single photon, as
shown in Fig.~\ref{fig:model}(b). For this system, continuum band
has a bandwidth of $4J_{\alpha }$ and their paired discrete levels,
denoted respectively by $E_{\alpha +}$ and $E_{\alpha -}$, are
gapped from either below or above, as illustrated in
Fig.~\ref{fig:model}(c). Reverted to the conventional language of
atomic scattering, the continua of eigenenergies can be considered
open channels of multiple admissible energy states in the continuous
range $\omega _{\alpha }-2J_{\alpha }<E<\omega _{\alpha }+2J_{\alpha
}$. Out of this range, the energy states can only admit two discrete
levels that associate with a non-real $k_{\alpha }$, representing
closed channels or bound states. These two discrete levels
\begin{equation}
E_{\alpha \pm }=\Omega \pm \frac{g_{\alpha }^{2}}{\sqrt{(E_{\alpha \pm
}-\omega _{\alpha })^{2}-4J_{\alpha }^{2}}}.  \label{eq:discrete_eigen}
\end{equation}%
are exactly solved from the above  discrete-coordinate scattering
equation. The dependence of these two bound state energies on the
various system parameters gives hints to their potential of
tunability and controllability. The readers familiar with the
approaches by Lee~\cite{lee54}, Fano~\cite{fano61}, and
Anderson~\cite{anderson61} shall also find our result here familiar.

The next logical step is to study how the resonance phenomenon
arises when two individual arrays are paired to form the H-shape
system. The scattering state $\left\vert \varphi \right\rangle $ of
the H-system for single-photon reads  $\left\vert \varphi
\right\rangle =\sum_{j}\left[ u_{\mathrm{A},g}(j)\left\vert
g,1_{j},0\right\rangle +u_{\mathrm{B},g}(j)\left\vert
g,0,1_{j}\right\rangle \right] +u_{e}\left\vert e,0,0\right\rangle
$. The probability amplitudes $u_{\mathrm{A},g}(j)$ and
$u_{\mathrm{B},g}(j)$ of the photonic occupation among the cavities
and $u_{e}$ of the atomic excitation in the system can still be
analyzed through the time-independent Schr\"{o}dinger equations,
leading to a pair of algebraic scattering equations similar to
Eq.~(\ref{eq:scatter_eqn}), one for the probability amplitudes in
each array. The distinction of the case here lies in the
adding $W_{\alpha }(E)=g_{\alpha }g_{\bar{\alpha}}u_{\bar{\alpha}%
,g}(0))/(E-\Omega )$ in the right hand side with $\alpha $ indexing
either A or B. We have used $\bar{\alpha}$ to indicate the dual
array relative to $\alpha $; namely, when $\alpha $ indexes A, then
$\bar{\alpha}$ indexes B and vice versa. This term reflects a
potential again contributed by the interaction of each array with
the central cavity. However, because of the coupling between the
central cavity and the dual array, additional contribution from the
dual array has to be considered.

\begin{figure}[tbp]
\includegraphics[bb=41bp 140bp 541bp 761bp,clip,width=8cm]{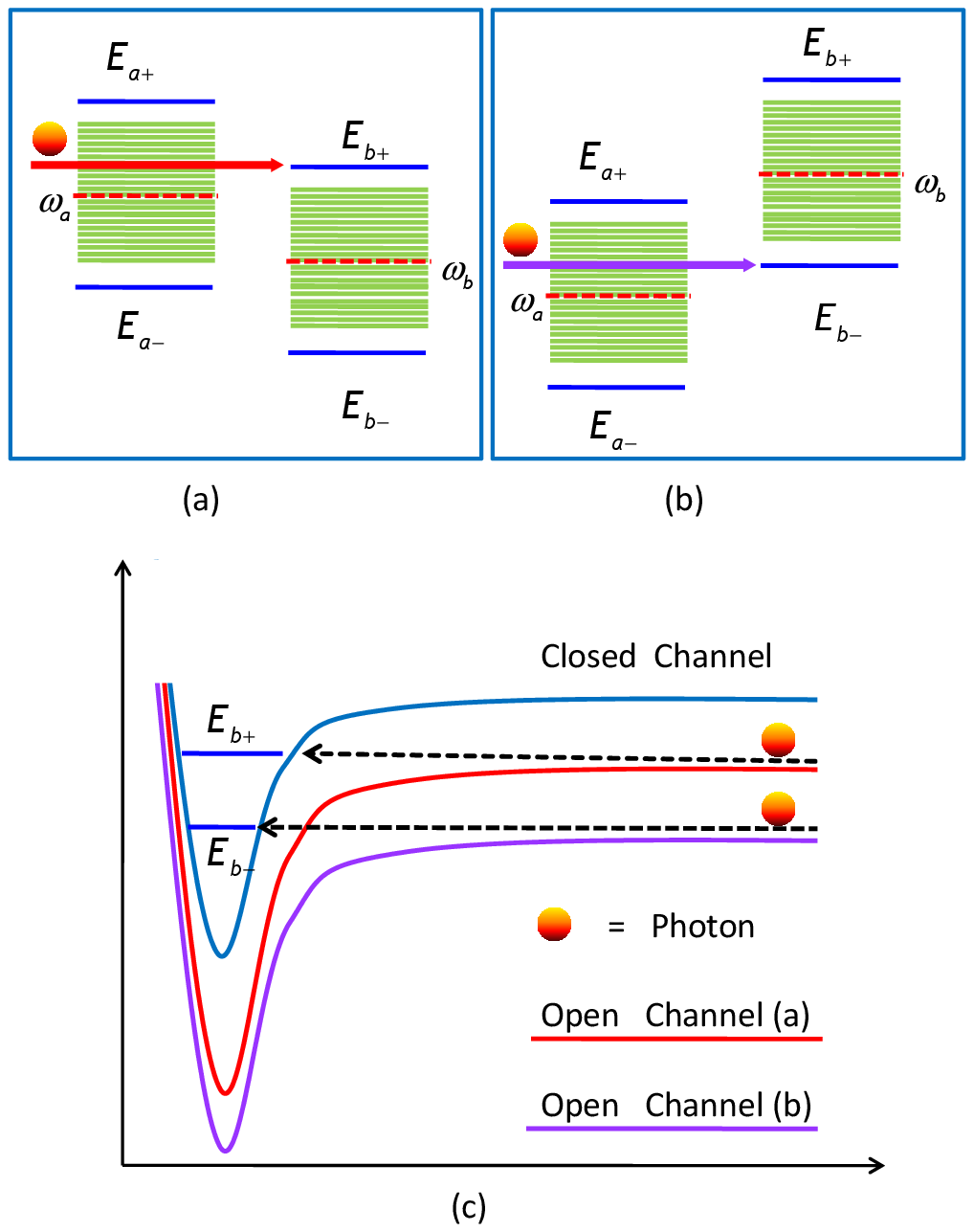}
\caption{Diagrams of the energy state distributions of the single
photon in array A and array B, showing the process of Feshbach
resonance between an open channel and a closed channel. Two
particular resonance cases exist for an incident photon inserted
into array A: (a) the photon energy level in the continuum band in
array A is resonant with the upper discrete level $E_{\mathrm{B}+}$
in array B; (b) the photon energy level in the continuum band in
array A is resonant with the lower discrete level $E_{\mathrm{B}-}$
in array B. (c) A profile plot illustrating the same two cases of
channel resonances relative to the cavity position $j$ the photon
occupies.} \label{fig:resonance}
\end{figure}

The set of solutions to the pair of scattering equations are many.
The portion we are concerned with are those illustrating a
simultaneously existent set of open channel and closed channel in
the two coupled resonator arrays. We select one of the particular
cases when a single photon is inserted into an open channel in array
A from the left for this purpose. The distribution of this single
photon in the array is then described by a plane wave,
$u_{\mathrm{A}}(j)=\exp (ikj)+r\exp (-ikj)\;(j<0);\;s\exp
(ikj)\;(j>0)$. $s$ and $r$ denote, respectively, the transmission
and the reflection coefficients of the optical plane wave,
indicating the scattering of photon by the effective potential
Eq.~(\ref{eq:eff_pot}) at the zeroth resonator in the
one-dimensional coordinate space. Meanwhile, the distribution
amplitude for the single photon in array B can be quasi-plane-wave
type, $u_{\mathrm{B}}(j)=C_{\mathrm{B}}\exp \left\{ -i\kappa
|j|\right\} $, with a complex wave number $\kappa $, and indicate a
closed channel, same as that of the individually discussed case.
These two distributions in the paired arrays, when combined through
the coupled scattering equations, give rise to a unified
dual-channel coupling equation
\begin{equation}
(1-s)\sin k\frac{J_{\mathrm{A}}}{g_{\mathrm{A}}}=C_{\mathrm{B}}\sin
\kappa
\frac{J_{\mathrm{B}}}{g_{\mathrm{B}}}=\frac{g_{\mathrm{A}}s+g_{\mathrm{B}}C_{\mathrm{B}}}{2i(\Omega
-E)}.  \label{eq:channel_eqn}
\end{equation}%
Therefore, the optical dual-channel resonance occurs when there
exists a solution of real $k$ and complex $\kappa $ to
Eq.~(\ref{eq:channel_eqn}) and the eigenenergy $E$ of the photon in
array A matches either of the discrete energy levels
$E_{\mathrm{B}\pm }$ of array B. The process is illustrated in
Fig.~\ref{fig:resonance} for two particular cases with $E$ matching
$E_{\mathrm{B}+}$ in Fig.~\ref{fig:resonance} (a) or
$E_{\mathrm{B}-}$ in Fig.~\ref{fig:resonance}(b). The two
possibilities of channel resonances is further illustrated in
Fig.~\ref{fig:resonance} (c) with the equivalent potential of array
A as a function of the resonator position $j$. The zeroth resonator
locates the position of local minimal energy for both the open
channels and the closed channel, reflecting the potential barrier
set up by the central cavity. The dual-channel resonance occurs as
well when the roles of array A and array B are exchanged.

\begin{figure}[tbp]
\subfigure[]{\includegraphics[width=8cm]{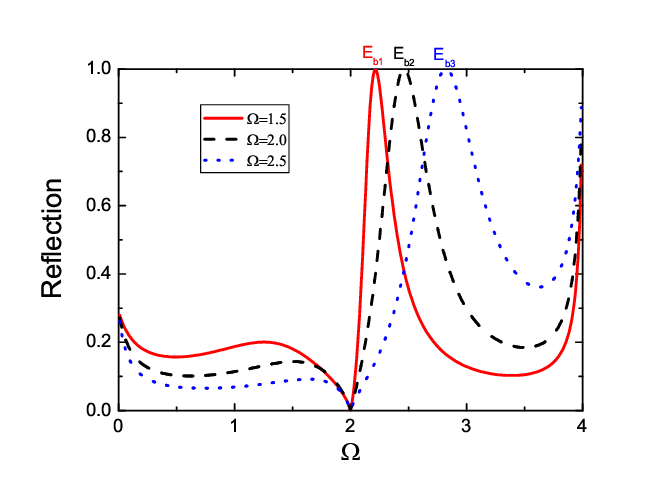}}\
\subfigure[]{\includegraphics[width=8cm]{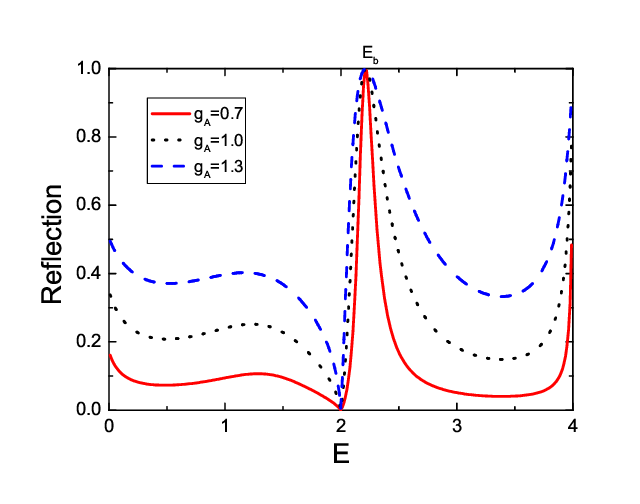}} \caption{Plots of the
norm-squared reflection coefficient $|1-s|^{2}$ against the
eigenenergy $E$ of a propagating photon in the coupled resonator
array A of the process illustrated in Fig.
\protect\ref{fig:resonance}(a). Two tuning parameters are varied:
(a) the level spacing $\Omega$ of the qubit controller; and (b) the
coupling coefficient $g_{\mathrm{A}}$ between the qubit controller
and the zeroth cavity in array A. Other parameters are chosen to be:
$J_{\mathrm{A}}=1$, $J_{\mathrm{B}}=0.5$,
$\protect\omega_{\mathrm{A}}=2$, $\protect\omega_{\mathrm{B}}=1$ and
$g_{\mathrm{B}}=0.7$. The incident energy ranges from $0$ to $4$ and
the continuum band for the array B is set $\left[0,2\right]$. The
corresponding bound state energies are marked as
$E_{b1}$,$E_{b2}$,$E_{b3}$ in (a) and $E_b$ in (b).}
\label{fig:reflec}
\end{figure}

The criteria of the dual-channel resonance can be met if the
transmission coefficient $s$ in Eq.~(\ref{eq:channel_eqn}) vanishes.
This condition implies the circumstance where the incident photon in
array A is totally reflected or scattered by the potential barrier
set up by the central cavity at position $j=0$. In other words, the
level spacing $\Omega $ of the controller becomes our tuning
parameter for the photonic  Feshbach resonance. Written in terms of
the other variables in the dual-channel coupling equation
$s=-2iC_{\mathrm{B}}J_{\mathrm{B}}(E-\Omega )\sin \kappa
/(g_{\mathrm{A}}g_{\mathrm{B}})-C_{\mathrm{B}}g_{\mathrm{B}}/g_{\mathrm{A}}$,
the transmission coefficient vanishes when $E=\Omega
-g_{\mathrm{B}}^{2}/(2iJ_{\mathrm{B}}\sin \kappa )$, leading to a
complex wave number $\kappa $ as expected. The complete reflection
in the open channel can be understood as the divergence effect of
s-wave scattering length for the usual Feshbach resonances in
three-dimensional space reduced to a version in one-dimensional
space. Eliminating the various variables, the transmission
coefficient can be expressed as a function of the incident energy
$E,i.e.,$

\begin{equation}
s=%
\begin{cases}
F_{\mathrm{A}}(E)\left[F_{\mathrm{A}}(E)-G_{-}(E)\right]^{-1}, & E>\omega_{%
\mathrm{B}}+2J_{\mathrm{B}} \\
F_{\mathrm{A}}(E)\left[F_{\mathrm{A}}(E)-G_{+}(E)\right]^{-1}, & E<\omega_{%
\mathrm{B}}+2J_{\mathrm{B}}%
\end{cases}
\label{eq:trans_coeff}
\end{equation}
where we have used the shorthands
$F_{\alpha}(E)=\sqrt{(E-\omega_{\alpha})^{2}-4J_{\alpha}^{2}}$ with
$\alpha\in\left\{ \mathrm{A},\mathrm{B}\right\} $ and
$G_{\pm}(E)=g_{\mathrm{A}}^{2}F_{\mathrm{B}}
(E)/\left((E-\Omega)F_{\mathrm{B}}(E)\pm g_{\mathrm{B}}^{2}\right)$.
The norm-squared reflection coefficient $|1-s|^{2}$ is plotted
against the photon energy in families of varying level spacing
$\Omega$ and coupling constant $g_{A}$ of the qubit controller in
Fig.~\ref{fig:reflec}. The photon encounters two kinds of
characteristic points while propagating through array A. The first
one is an indifferentiable turning point where $s=1$ or
$E=\omega_{\mathrm{B}}+2J_{\mathrm{B}}$. The potential barrier
becomes transparent and the photon is completely transmitted because
of the matching coupling between the qubit controller and the dual
array B. The second one is the maximum point where the photon is
fully reflected when the transmission coefficient is vanishing
$s=0$. We hence see the shifting of this peak while $\Omega$ is
varied. The reliance on the coupling coefficient $g_{\mathrm{A}}$
determines the width of the peaking.

\begin{figure}[tbp]
\subfigure[]{\includegraphics[width=8cm]{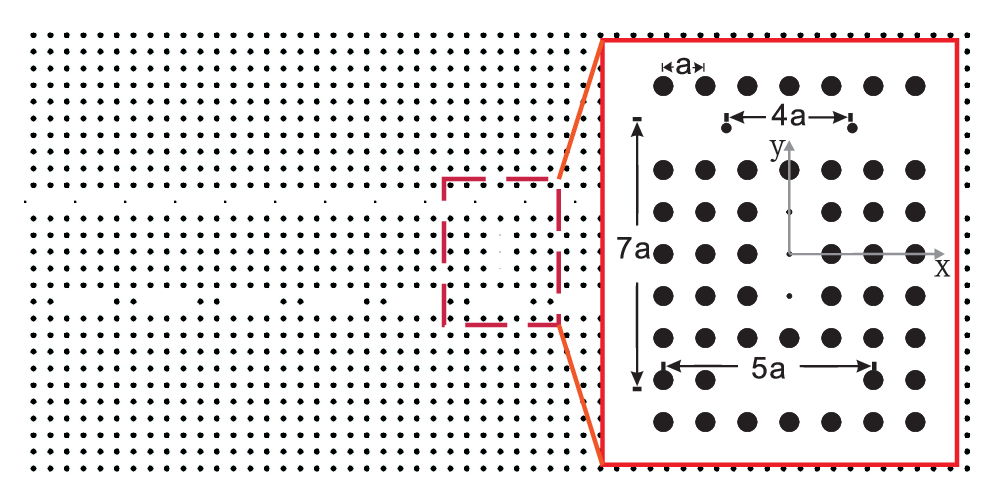}}
\subfigure[]{\includegraphics[width=8cm]{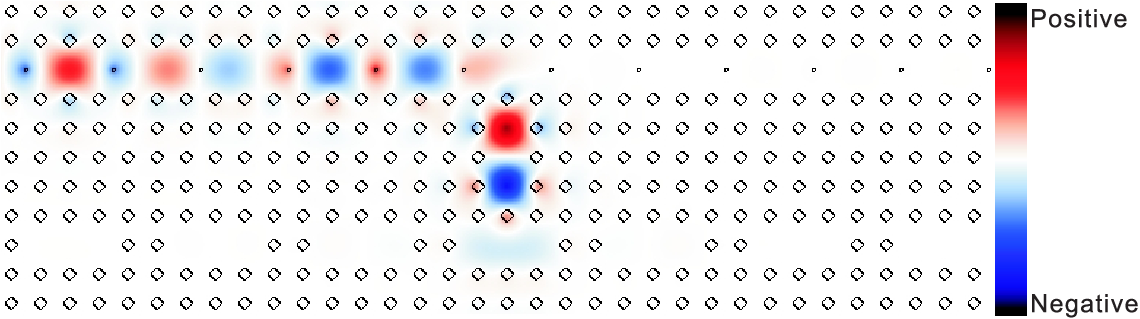}}
\caption{An experimental protocol based on a photonic crystal made
up of silicon rods of radius $0.2a$. (a) The structure of the
design: the upper waveguide is implemented by removing a row of
original rods and substituting with a set of rods of radius $0.1a$
and spacing $d_{\mathrm{UP}}=4a$. The lower waveguide, $D=7a$ apart
from the upper one, is constructed by removing three rods out of
every five rods, i.e. lattice spacing $d_{\mathrm{DOWN}}=5a $. The
central cavity is created by reducing the radius of three
vertically-placed rods between the two waveguides to $0.05a$. (b)
Plotting the steady electric field vector for an incident wave of
frequency $\omega_0=0.3628\times2\protect\pi c/a$ with
TM-polarization.} \label{fig:pc}
\end{figure}

Next we numerically examine the feasibility of our theoretical
prediction on a two-dimensional photonic
crystal~\cite{fanapl02,joan07}. The crystal is made up of a square
lattice of high-index dielectric rods of radius $0.2a$, $0.1a$ and
$0.05a$, where $a$ is the lattice spacing. The artificial design is
made by two parallel waveguides of coupled defected cavity arrays
linked through a central defected cavity on the two-dimensional
photonic crystal, as illustrated in Fig.~\ref{fig:pc}(a). The two
resonator arrays~\cite{yariv99} is constructed with different
frequencies, inter-resonator tunneling rates, and coupling strengths
with the central cavity.

For this photonic crystal, the material of all the rods is assumed
to be silicon, with a dielectric constant $\epsilon =11.56$, and the
background is filled by air. We make the simulation of the designed
structure with the finite-difference time-domain (FDTD)
method~\cite{FDTD} in freely available Meep code~\cite{Meep}. The
steady field vector of the incident wave at
frequency $\omega _{0}=0.3628\times 2\pi c/a$ is plotted in Fig. \ref{fig:pc}%
(b), with red showing positive amplitudes pointing out from the
plane and blue negative amplitudes into the plane. The wave travels
horizontally from left to right and hence, according to the
convention for characterizing photonic crystals, carries a
transverse-magnetic (TM) polarization. The notice-worthy region is
located at the center where the highly-saturated colors indicate a
localized bounded photon from the lower waveguide. Moreover, the
blank portion in the upper waveguide indicates a completely
reflected wave. Finally, we point out that, though the numerical
simulation based on FDTD is of classical, but the weak light
calculation can also reflect the single photon nature with the
intensity distribution illustrated in Fig.~\ref{fig:pc}(b), which is
only relevant to the first order coherence function.

In conclusion, we have shown the existence of a photonic bound
state in a qubit-controlled coupled resonator array and predicted
photonic  Feshbach resonance emerges from a pair of these coupled
resonator arrays coupled in an H-shape fashion. An FDTD simulation
of the system implemented on a photonic crystal has verified the
validity of the proposal. The resonance phenomenon arises from the
dual-channel coupling between an unbound state in one array and a
bound state in the other, the occurring moment of which is
indicated by a total reflection of an incident photon in the
array. Our analysis for the resonant scattering process was
carried out for the single photon case and did not rely on the
photonic statistics. Our prediction here is thus applicable to
fermionic models, such as the electron transportation along a
H-shape array of quantum dots. For the case where multiple photons
are assumed to exist in the arrays, Bethe-ansatz must be used for
the analysis and we shall defer its discussion in a future work.

\begin{acknowledgments}
C.P.S. acknowledges the helpful discussion with S. Yang, Peng Zhang
and X. H. Wang. This work is supported by NSFC No.10474104, No.
60433050, and No. 10704023, NFRPC No. 2006CB921205 and 2005CB724508.
\end{acknowledgments}

\end{document}